\documentclass[10pt]{article}
\usepackage[margin=1in]{geometry} 
\usepackage{amsmath,amsthm,amssymb, physics, mathtools, subfig, float}

\usepackage[numbers,sort&compress]{natbib}

\usepackage{authblk}
\usepackage{hyperref}
\newcommand*{\email}[1]{\normalsize\href{mailto:#1}{#1}}
\def\defeq{\mathrel{\mathop:}=}

\newcommand{\vect}[1]{\boldsymbol{#1}}

\begin{document}

\title{Hamiltonian Dynamics of Semiclassical Gaussian Wave Packets in Electromagnetic Potentials}
\author{Nolan King}
\author{Tomoki Ohsawa}
\affil{Department of Mathematical Sciences\\
  The University of Texas at Dallas\\
  800 W Campbell Rd, Richardson, TX 75080-3021, USA\\
  \email{Nolan.King@utdallas.edu}, \email{tomoki@utdallas.edu}
}



\date{\vspace{-5ex}}
\maketitle

\begin{abstract}
 We extend our previous work on symplectic semiclassical Gaussian wave packet dynamics to incorporate electromagnetic interactions by including a vector potential. The main advantage of our formulation is that the equations of motion derived are naturally Hamiltonian. We obtain an asymptotic expansion of our equations in terms of $\hbar$ and show that our equations have $\mathcal{O}(\hbar)$ corrections to those presented by Zhou, whereas ours also recover the equations of Zhou in the case of a linear vector potential and quadratic scalar potential. One and two dimensional examples of a particle in a magnetic field are given and numerical solutions are presented and compared with the classical solutions and the expectation values of the corresponding observables as calculated by the Egorov or Initial Value Representation (IVR) method. We numerically demonstrate that the $\mathcal{O}(\hbar)$ correction terms improve the accuracy of the classical or Zhou's equations for short times in the sense that our solutions converge to the expectation values calculated using the Egorov/IVR method faster than the classical solutions or those of Zhou as $\hbar \to 0$.
\end{abstract}


\section{Introduction}
\subsection{Motivation}
Gaussian wave packets have historically been used to solve the time-dependent semiclassical Schr\"odinger equation~\cite{He1975a,He1976b,He1981,Ha1980,Ha1998,Li1986}. While the Schr\"odinger equation is computationally non-trivial to solve in the semiclassical regime, those methods using the Gaussian wave packets provide an alternative set of differential equations that may be solved for the time-dependent parameters of the Gaussian wave packet. The Gaussian wave packet is an ansatz for an exact solution in the case of linear vector potentials with quadratic scalar potentials (see \citet{Ha1998}), and gives a good short time approximation of the solution for other potentials in the semiclassical regime as shown by \citet{Zh2014}.

However, the set of differential equations of Zhou for the parameters is not a Hamiltonian system in general.
Given that the equations of motion for a classical particle is a Hamiltonian system and also that the Schr\"odinger equation is a (infinite-dimensional) Hamiltonian system as we will explain in a moment, it is rather natural to seek a Hamiltonian formulation of the dynamics of the Gaussian wave packet.
This was the main motivation of our previous work~\cite{OhLe2013} on the symplectic/Hamiltonian formulation of the Gaussian wave packet dynamics.

In this paper, we utilize the same symplectic-geometric framework to derive a Hamiltonian system of equations governing the evolution the wave packet parameters under the influence of electromagnetic fields by taking into account a vector potential.
Semiclassical dynamics under the influence of electromagnetic fields has been of great interest recently because of its significance in quantum control and
solid state physics.

\subsection{Hamiltonian Formulation of Classical Dynamics}
It is well known that the equations of motion of a classical particle in $\mathbb{R}^{d}$ is a Hamiltonian system.
From the symplectic-geometric point of view, one takes the cotangent bundle $T^{*}\mathbb{R}^{d} \cong \mathbb{R}^{2d} = \{ (q, p) \,|\, q, p \in \mathbb{R}^{d} \}$ as the phase space and define the classical symplectic form $\Omega_{0} \defeq \mathbf{d}q_{i} \wedge \mathbf{d}p_{i}$ (the Einstein summation convention is assumed).
This renders $T^{*}\mathbb{R}^{d}$ a symplectic manifold.
We also define a Hamiltonian function $H_{0}\colon T^{*}\mathbb{R}^{d} \to \mathbb{R}$ as
\begin{equation}
  \label{eq:H_0}
  H_{0}(q,p) \defeq \frac{1}{2m}(p - \vect{A}(q))^{2} + V(q),
\end{equation}
where $V\colon \mathbb{R}^d \rightarrow \mathbb{R}$ and $\vect{A}\colon \mathbb{R}^d \rightarrow \mathbb{R}^d$ are scalar and vector potentials respectively, and we set the charge to be 1 for simplicity.

Now let $X_{H_{0}} = \dot{q}_{i} \frac{\partial}{\partial q_{i}} + \dot{p}_{i} \frac{\partial}{\partial p_{i}}$ be the vector field on $T^{*}\mathbb{R}^{d}$ defined by $\textbf{i}_{X_{H_{0}}}\Omega_{0} =  \textbf{d}H_{0}$ where $\mathbf{i}$ stands for the contraction.
Then the equation yields the equations of motion of the classical particle in the electromagnetic field:
\begin{equation}
  \label{eq:classical}
  \dot{q} = \frac{1}{m}(p - \vect{A}(q)),
  \qquad
  \dot{p} = -\frac{1}{2m} \grad_{q}\left( |\vect{A}(q)|^{2} - 2\vect{A}(q) \cdot p \right) - \grad V(q),
\end{equation}
where $\grad_{q}$ stands for the gradient with respect to the variable $q$, and $|\,\cdot\,|$ stands for the Euclidean distance in $\mathbb{R}^{d}$.

\subsection{Hamiltonian Formulation of the Schr\"odinger Equation}
We may generalize the notion of Hamiltonian system as follows:
Let $P$ be a symplectic manifold, i.e., a manifold equipped with a closed non-degenerate 2-form $\Omega$, and let $H\colon P \to \mathbb{R}$ be a smooth function.
Then we define the \textit{Hamiltonian vector field} $X_{H}$ on $P$ corresponding to the Hamiltonian function $H$ by setting $\textbf{i}_{X_H}\Omega =  \textbf{d}H$.
The vector field $X_{H}$ then defines the evolution equation on $P$.
We may take it as the definition of a generalized Hamiltonian system; see, e.g., \citet{MaRa1999} for more details.

We may now formulate the time-dependent Schr\"odinger equation as a Hamiltonian system as follows:
Let $\mathcal{H} \defeq L^2(\mathbb{R}^d)$ with the standard (right-linear) inner product $\left\langle\,\cdot\,,\,\cdot\,\right\rangle$, and equip it with the symplectic form $\Omega_{\mathcal{H}}(\psi_1, \psi_2)  \defeq 2 \hbar \Im \left\langle\psi_1, \psi_2\right\rangle$, and take the expectation value $\langle H\rangle\colon \mathcal{H} \to \mathbb{R}$ of the Hamiltonian operator as the Hamiltonian function. Then the Hamiltonian vector field $X_{\langle H\rangle}$ on $\mathcal{H}$ defined by $\textbf{i}_{X_{\langle H\rangle}}\Omega_{\mathcal{H}} =  \textbf{d}\langle H\rangle$ yields the usual time-dependent Schr\"odinger equation
\begin{equation}
  \label{eq:Schroedinger}
  \text{i}\hbar \frac{\partial}{\partial t} \psi = \hat{H} \psi,
\end{equation}
where $\hat{H}$ is the Hamiltonian operator defined below.

\subsection{Geometry of Reduced Models}
Given that the basic equations of classical and quantum dynamics are both Hamiltonian systems, it is natural to expect that the basic equations of semiclassical dynamics---or more generally approximation/reduced models of quantum dynamics---are Hamiltonian as well.
Is there a way to exploit the above symplectic structure $\Omega_{\mathcal{H}}$ on $\mathcal{H} = L^2(\mathbb{R}^d)$ to find a Hamiltonian formulation of reduced models?

\begin{figure}[ht]
  \centering
  \includegraphics[width=.65\linewidth]{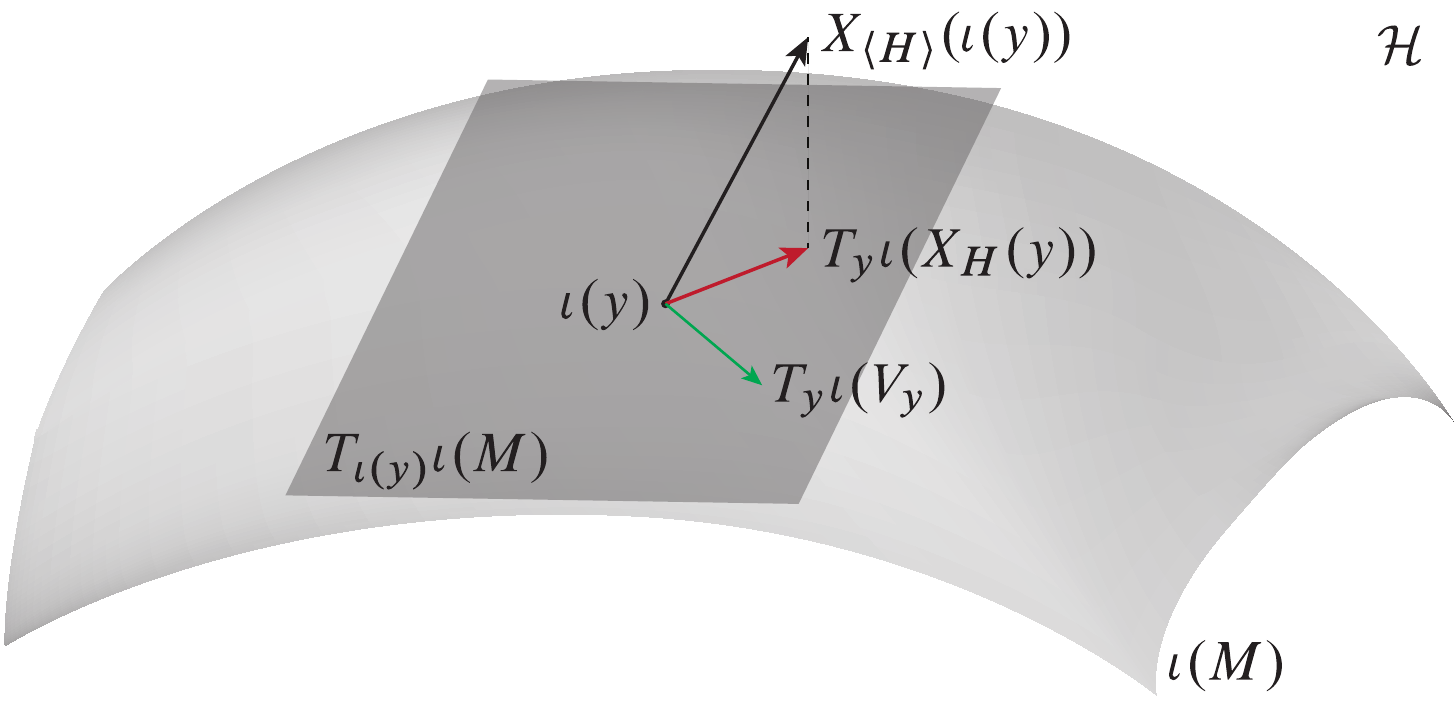}
  \caption{The Hamiltonian Vector Field $X_H$ gives the best approximation on $M$ of the vector field $X_{\langle H\rangle}$.}
  \label{fig:BestApproximation}
\end{figure}

\citet{Lu2008} (see also \citet{KrSa1981}) came up with a general prescription to achieve this by geometrically interpreting approximation models of quantum dynamics.
Suppose that we have an ansatz $\varphi\colon M \to \mathcal{H} \defeq L^{2}(\mathbb{R}^{d}); y \mapsto \varphi(y;\,\cdot\,)$ for the solution of the Schr\"odinger equation, where $M$ is a finite-dimensional manifold (where the parameters for the ansatz live).
The parameters $y$ evolve in time according to the dynamics in $M$ to be determined, and the time evolution $t \mapsto \varphi(y(t);\,\cdot\,)$ gives an approximation to the solution of the Schr\"odinger equation~\eqref{eq:Schroedinger}.
\citet{Lu2008} (see also \citet{OhLe2013}) showed that one can achieve the best approximation in $M$ as follows:
Consider the embedding $\iota\colon M \rightarrow \mathcal{H}$ defined by the ansatz $\varphi$ as $\iota(y) \defeq \varphi(y; \,\cdot\,)$.
Then we can pull back the symplectic form $\Omega_{\mathcal{H}}$ to $M$ to obtain a 2-form $\Omega \defeq \iota^{*}\Omega_{\mathcal{H}}$ on $M$.
Under a certain technical condition (see \cite{Lu2008} and \cite[Proposition~2.1]{OhLe2013} for details), $\Omega$ defines a symplectic form on $M$.
One may also define the pull-back $H \defeq \langle H\rangle \circ \iota$ of the Hamiltonian function, i.e., $H(y) \defeq \left\langle \varphi(y;\,\cdot\,), \hat{H} \varphi(y;\,\cdot\,)\right\rangle$.
Then we may define the Hamiltonian vector field $X_H$ on $M$ by setting $\textbf{i}_{X_H}\Omega =  \textbf{d}H$.
\citet{Lu2008} showed that this gives the least squares approximation of the vector field $X_{\langle H\rangle}$ in the following sense: For any $y \in M$ and any $V_{y} \in T_{y}M$,
\begin{equation*}
  \| X_{\langle H\rangle}(\iota(y)) - T_{y}\iota(X_H(y)) \|
  \le
  \| X_{\langle H\rangle}(\iota(y)) - T_{y}\iota(V_{y}) \|
\end{equation*}
in terms of the $L^{2}$ norm in $\mathcal{H} = L^{2}(\mathbb{R}^{d})$; see Fig.~\ref{fig:BestApproximation}.

\section{Hamiltonian Dynamics of Gaussian Wave Packets in Electromagnetic Potentials}
\subsection{Gaussian Wave Packet in Electromagnetic Potentials}
Our ansatz or approximation/reduced model is the Gaussian wave packet of \citet{He1975a,He1976b,He1981} and \citet{Ha1980,Ha1998} (see also \citet{Li1986}):
\begin{equation}
  \label{eq:GWP}
  \chi_M({q}, {p}, \mathcal{A}, \mathcal{B}, \phi, \delta; x) \defeq \exp{\frac{\text{i}}{\hbar} \left( \frac{1}{2} (x-q)^T (\mathcal{A} + \text{i} \mathcal{B})(x-q) + p\cdot (x-q) + \phi + \text{i} \delta \right) },
\end{equation}
where $(q, p) \in T^{*}\mathbb{R}^d \cong \mathbb{R}^{2d}$ is the phase space center, $\phi$ $\in$ $\mathbb{S}^1$ is the phase factor, $\delta$ $\in$ $\mathbb{R}$ controls the norm, and $\mathcal{A} + \text{i} \mathcal{B} \in \Sigma_d \defeq \{ \mathcal{A} + \text{i} \mathcal{B} \in \mathbb{C}^{d\times d} \,|\, \mathcal{A},\mathcal{B} \in \text{Sym}_d(\mathbb{R}), \mathcal{B}-\text{positive definite}\}$.
Note that the above Gaussian is not normalized:
\begin{equation}
  \label{eq:N}
  \mathcal{N}(\mathcal{B},\delta) \defeq \norm{\chi(y;\,\cdot\,)}^{2}
  = \sqrt{ \frac{(\pi\hbar)^{d}}{\det \mathcal{B}} }\, \exp\left( -\frac{2\delta}{\hbar} \right),
\end{equation}
where we set $y = ({q}, {p}, \mathcal{A}, \mathcal{B}, \phi, \delta)$.
We will address this issue later.

Following the geometric picture of \citet{Lu2008} described above, we define $M$ to be the space of the above parameters:
\begin{equation*}
  M \defeq T^*\mathbb{R}^d \times \Sigma_d \times \mathbb{S}^1 \times \mathbb{R},
\end{equation*}
and consider the embedding $\iota\colon M \to \mathcal{H}$ defined as $\iota(y) \defeq \chi_{M}(y;\,\cdot\,)$.
Then one can show that the pull-back $\Omega_{M} \defeq \iota^{*}\Omega_{\mathcal{H}}$ is in fact a symplectic form on $M$; see \citet[Section~3]{OhLe2013}.

In this paper, we would like to incorporate the effect of electromagnetic fields to the dynamics of the Gaussian wave packet.
So we take Hamiltonian operator $\hat{H}$ to be 
\begin{equation*}
  \hat{H} \defeq \frac{1}{2m}\bigg( -\text{i}\hbar \grad - \vect{A}(x) \bigg)^2 + V(x),
\end{equation*}
where we assume that the scalar and vector potentials $V\colon \mathbb{R}^d \rightarrow \mathbb{R}$ and $\vect{A}\colon \mathbb{R}^d \rightarrow \mathbb{R}^d$ are smooth functions; we write the $i$-th component of $\vect{A}_{i}$ as opposed to the more conventional $A_{i}$ in order to make it more conspicuous and to avoid possible confusions with the components of $\mathcal{A}$.

One may then evaluate the pull-back $H_{M} \defeq \langle H\rangle \circ \iota\colon M \to \mathbb{R}$ of the Hamiltonian $\langle H\rangle$ by evaluating the expectation value of the Hamiltonian operator as follows:
\begin{equation}
  \label{eq:H_M}
  \begin{split}
    H_{M}(y) \defeq \langle H\rangle \circ \iota(y)
    &= \left\langle \chi_{M}(y;\,\cdot\,), \hat{H}\chi_{M}(y;\,\cdot\,)\right\rangle \\
    &= \mathcal{N}(\mathcal{B},\delta)\biggl(
    \frac{p^2}{2m} + \frac{\hbar}{4m} \Tr( \mathcal{B}^{-1}(\mathcal{A}^2 + \mathcal{B}^2) ) - \frac{1}{m} \left\langle \vect{A}(x) \cdot p\right\rangle \\
    &\quad+ \frac{\hbar}{2m}\left\langle \Tr( D\vect{A}^T(x) \mathcal{A}\mathcal{B}^{-1})\right\rangle + \frac{1}{2m} \left\langle | \vect{A}(x) |^2 \right\rangle + \left\langle V(x) \right\rangle
    \biggr),
  \end{split}
\end{equation}
where $\vect{A}$ is regarded as a column vector, $D\vect{A}(x)$ is the $d \times d$ matrix whose $(i,j)$-component is $\frac{\partial \vect{A}_i}{\partial x_j}(x)$, and $\langle\,\cdot\,\rangle$ stands for the expectation value of an observable with respect to the \textit{normalized} Gaussian $\chi_{M}/\norm{\chi_{M}}$:
For any smooth function $\mathcal{U}\colon \mathbb{R}^{d} \to \mathbb{R}$ satisfying a certain growth condition (see Section~\ref{ssec:asymptotic_expansion}),
\begin{equation}
  \label{eq:exval}
  \begin{split}
    \left\langle \mathcal{U}(x) \right\rangle
    &\defeq \frac{1}{\mathcal{N}(\mathcal{B},\delta)} \left\langle \chi_{M}(y;\,\cdot\,), \mathcal{U}(\,\cdot\,) \chi_{M}(y;\,\cdot\,) \right\rangle \\
    &= \frac{1}{\mathcal{N}(\mathcal{B},\delta)} \int_{\mathbb{R}^d} \mathcal{U}(x) \exp\left( -\frac{1}{\hbar} (x-q)^T \mathcal{B} (x-q) \right) \text{d}x.
  \end{split}
\end{equation}

\subsection{Hamiltonian Formulation of Gaussian Wave Packet Dynamics}
One may now certainly formulate a Hamiltonian system on $M$ using the above pull-backs of the symplectic form and the Hamiltonian.
However, the pull-back of the symplectic form turns out to be very cumbersome; neither does it provide much insight into its relationship with the symplectic form $\Omega_{0}$ of classical dynamics; see \citet[Eq.~(10)]{OhLe2013}.

Fortunately, there is a way around it to obtain a simpler and more appealing formulation by exploiting the inherent symmetry of the system~\cite[Section~4]{OhLe2013}.
First observe that the Hamiltonian~\eqref{eq:H_M} does not depend on the phase factor variable $\phi$; that is, the Hamiltonian is invariant under the following $\mathbb{S}^{1}$-action on the manifold $M$:
\begin{equation*}
  \mathbb{S}^{1} \times M \to M;
  \quad
  \left( \theta, (q, p, \mathcal{A}, \mathcal{B}, \phi, \delta) \right) \mapsto (a, p, \mathcal{A}, \mathcal{B}, \phi + \hbar\,\theta, \delta).
\end{equation*}
This action turns out to be symplectic and the corresponding momentum map (Noether conserved quantity) is
\begin{equation*}
  \mathbf{N}\colon M \to \mathbb{R};
  \qquad
  \mathbf{N}(y) \defeq -\hbar\,\mathcal{N}(\mathcal{B},\delta).
\end{equation*}
It is natural to look at the level set $\mathbf{N}^{-1}(-\hbar)$ because, in view of the definition~\eqref{eq:N} of $\mathcal{N}$, this level set corresponds to the choice of the parameter $\delta$ so that the Gaussian $\chi_{M}$ is normalized, i.e., $\norm{\chi(y;\,\cdot\,)} = 1$.
Furthermore, one may eliminate the variables $(\phi,\delta)$ from the formulation, because now we may apply the Marsden--Weinstein reduction~\cite{MaWe1974} (see also \citet[Sections~1.1 and 1.2]{MaMiOrPeRa2007}) to obtain the reduced symplectic manifold
\begin{equation*}
  \overline{M} \defeq \mathbf{N}^{-1}(-\hbar)/\mathbb{S}^{1} = T^{*}\mathbb{R}^{d} \times \Sigma_{d}.
\end{equation*}
See \cite[Section~4]{OhLe2013} for the details of this reduction.

As a result, the symplectic form $\Omega_{M}$ on $M$ gives rise to the following symplectic form $\Omega_{\hbar}$ on $\overline{M}$:
\begin{equation}
  \label{eq:Omega}
  \begin{split}
    \Omega_{\hbar} &= \textbf{d}q_i \wedge \textbf{d}p_i + \frac{\hbar}{4} \mathcal{B}^{-1}_{ik} \mathcal{B}^{-1}_{lj}\textbf{d}\mathcal{A}_{ij}\wedge \textbf{d}\mathcal{B}_{kl}\\
    &= \textbf{d}q_i \wedge \textbf{d}p_i + \frac{\hbar}{4} \textbf{d}\mathcal{B}^{-1}_{ij} \wedge \textbf{d}\mathcal{A}_{ij}.
  \end{split}
\end{equation}
Notice that the symplectic form is very simple and also appealing because it has an additional $\mathcal{O}(\hbar)$ correction term compared to the classical symplectic form $\Omega_{0}$.
It is also clear from the above expressions that $\mathcal{B}^{-1}$ and $\mathcal{A}$ are conjugate variables.
The corresponding Poisson Bracket is then
\begin{align*}
  \left\{ F, G \right\}_{\hbar} \defeq
  \frac{\partial F}{\partial q_i}\frac{\partial G}{\partial p_i} - \frac{\partial G}{\partial q_i}\frac{\partial F}{\partial p_i}
  + \frac{4}{\hbar}\left(
  \frac{\partial F}{\partial \mathcal{B}_{jk}^{-1}}\frac{\partial G}{\partial \mathcal{A}_{jk}} - \frac{\partial G}{\partial \mathcal{B}^{-1}_{jk}}\frac{\partial F}{\partial \mathcal{A}_{jk}}
  \right).
\end{align*}

Since we are now looking at the normalized Gaussian, we have $\mathcal{N}(\mathcal{B},\delta) = 1$, and thus the reduced Hamiltonian $H\colon \overline{M} \to \mathbb{R}$ becomes
\begin{equation}
  \label{eq:H}
  \begin{split}
    H(q, p, \mathcal{A}, \mathcal{B})
    &= \frac{p^2}{2m} + \frac{\hbar}{4m} \Tr( \mathcal{B}^{-1}(\mathcal{A}^2 + \mathcal{B}^2) ) - \frac{1}{m} \left\langle \vect{A}(x) \cdot p\right\rangle \\
    &\quad+ \frac{\hbar}{2m}\left\langle \Tr( D\vect{A}^T(x) \mathcal{A}\mathcal{B}^{-1})\right\rangle + \frac{1}{2m} \left\langle | \vect{A}(x) |^2 \right\rangle + \left\langle V(x) \right\rangle.
  \end{split}
\end{equation}
The Hamiltonian vector field $X_{H}$ on $\overline{M}$ defined by the Hamiltonian system $\textbf{i}_{X_{H}} \Omega_{\hbar} = \textbf{d}H$ or equivalently $\dot{\bar{y}} = \left\{\bar{y}, H\right\}_{\hbar}$ with $\bar{y} = (q, p, \mathcal{A}, \mathcal{B})$ gives the following set of ordinary differential equations:
\begin{equation}
  \label{eq:semiclassical-original}
  \begin{split}
    \dot{q}_i &= \frac{1}{m} \left( p_i - \left\langle \vect{A}_i(x) \right\rangle \right), \\
    \dot{p}_i &= - \frac{1}{2m} \left(
      \left\langle D_i | \vect{A}(x) |^2  \right\rangle
      - 2 \left\langle D_i \vect{A}_j(x) p_j \right\rangle
    \right)
     - \frac{\hbar}{2m}\left\langle \vect{A}_k(x)  \mathcal{A}_{kj}\mathcal{B}_{ji}^{-1}  \right\rangle  - \left\langle D_i V(x)\right\rangle,\\
    \dot{\mathcal{A}}_{ij} &= -\frac{1}{m} ( \mathcal{A}^2 - \mathcal{B}^2)_{ij}  + \frac{1}{m} \left\langle D^2_{ij} \vect{A}_k(x) p_k - D_k \vect{A}_i(x) \mathcal{A}_{kj} - \mathcal{A}_{ik} D_j \vect{A}_k(x) - \frac{1}{2} \left\langle D^2_{ij} | \vect{A}(x) |^2\right\rangle \right\rangle, \\
    &\quad - \frac{\hbar}{2m}\left\langle D^2_{ij} (D_l \vect{A}_k(x) \mathcal{A}_{lm} \mathcal{B}_{mk}^{-1})\right\rangle - \left\langle D^2_{ij}  V(x) \right\rangle,\\
    \dot{\mathcal{B}}_{ij} &=  - \frac{1}{m} (\mathcal{A}\mathcal{B} + \mathcal{B} \mathcal{A})_{ij} + \frac{1}{ m} \left\langle \mathcal{B}_{ik} D_j \vect{A}_k(x) + D_k \vect{A}_i(x) \mathcal{B}_{kj} \right\rangle,
  \end{split}
\end{equation}
where $(D \vect{A})_{ij} = D_j \vect{A}_i = \frac{\partial \vect{A}_i}{\partial x_j}$, and $D^2_{ij} \vect{A}_k = \frac{\partial  \vect{A}_k}{\partial x_i \partial x_j} $.

\section{Asymptotic Expansion}
\subsection{Asymptotic Expansion of Hamiltonian}
\label{ssec:asymptotic_expansion}
While the above set~\eqref{eq:semiclassical-original} of equations is Hamiltonian by construction, it has the drawback that it is not in a closed form: The potential terms---involving either the scalar potential $V$ or the vector potential $\vect{A}$---appear as expectation values (with respect to the normalized Gaussian).
Unfortunately, it is impossible to explicitly evaluate these expectation values unless $V$ and $\vect{A}$ are polynomials.

So we apply Laplace's method to obtain an asymptotic expansions of the integrals as $\hbar \to 0$ (see, e.g., \citet[Chapter~3]{Mi2006}).
Assuming that the Gaussian is normalized, i.e., $\mathcal{N}(\mathcal{B},\delta) = 1$, each potential term is of the form (see \eqref{eq:exval}):
\begin{equation*}
  \left\langle \mathcal{U} \right\rangle(q,\mathcal{B}) = \int_{\mathbb{R}^d} \mathcal{U}(x) \exp\left( -\frac{1}{\hbar} (x-q)^T \mathcal{B} (x-q) \right) \text{ d}x,
\end{equation*}
As is proved in \citet[Proposition~7.1]{OhLe2013} (see also \citet[Section 3.7]{Mi2006}), if $\mathcal{U}$ satisfies a certain growth condition as $|x| \to \infty$, then $\left\langle \mathcal{U}\right\rangle$ has the following asymptotic expansion:
\begin{equation}
  \label{eq:asymptotic_expansion}
  \left\langle\mathcal{U}\right\rangle(q,\mathcal{B}) = \mathcal{U}(q) + \frac{\hbar}{4} \Tr( \mathcal{B}^{-1} D^2 \mathcal{U}(q))  + \mathcal{O}(\hbar^2) \quad \text{as} \quad \hbar \rightarrow 0,
\end{equation}
where $D^2 \mathcal{U}(q)$ is the Hessian matrix of $\mathcal{U}(x)$ evaluated at $x = q$.
We note in passing that this asymptotic expansion is exact (i.e., the $O(\hbar^{2})$ term vanishes) if $\mathcal{U}$ is quadratic.

As a result, we have the following asymptotic expansion for the Hamiltonian $H$ from \eqref{eq:H_M}:
\begin{equation*}
  H = H_{\hbar} + \mathcal{O}(\hbar^{2}) \quad \text{as} \quad \hbar \rightarrow 0
\end{equation*}
with
\begin{equation}
  \label{eq:H_hbar}
  \begin{split}
    H_{\hbar}(q, p, \mathcal{A}, \mathcal{B}) &\defeq \frac{1}{2m}(p - \vect{A}(q))^{2} \\
    &\quad+ \frac{\hbar}{4m} \Tr\left(
      \mathcal{B}^{-1} \left(
        \mathcal{A}^2 + \mathcal{B}^2 - D \vect{A}^T(q) \mathcal{A}  - \mathcal{A} D \vect{A}(q) - D^2 (\vect{A}(q) \cdot p) + \frac{1}{2}  D^2 |\vect{A}(q)|^2
      \right)
    \right)\\
    &\quad+ V(q) + \frac{\hbar}{4} \Tr (\mathcal{B}^{-1} D^2 V(q) ).
  \end{split}
\end{equation}
Notice that, just as for the symplectic form $\Omega$ in \eqref{eq:Omega}, this semiclassical Hamiltonian $H$ has an additional $\mathcal{O}(\hbar)$ correction term compared to the classical Hamiltonian $H_{0}$ from \eqref{eq:H_0}.

One may now replace the Hamiltonian $H$ by $H_{\hbar}$ to define the Hamiltonian vector field $X_{H_{\hbar}}$ as $\textbf{i}_{X_{H_{\hbar}}} \Omega_{\hbar} = \textbf{d}H_{\hbar}$.
Then the vector field $X_{H_{\hbar}}$ yields
\begin{equation}
  \label{eq:semiclassical}
  \begin{split}
    \dot{q}_i &= \frac{p_i}{m} -\frac{\vect{A}_i(q)}{m} - \frac{\hbar}{4m} \mathcal{B}_{jk}^{-1} D^2_{kj} \vect{A}_i(q), \\
    \dot{p}_i &=  - \frac{1}{2m}D_i \left(
      |\vect{A}(q)|^2 + \frac{\hbar}{4} ( \mathcal{B}_{jk}^{-1} D_{kj}^2 |\vect{A}(q)|^2 )
      -2\vect{A}_k(q) p_k  - \frac{\hbar}{2}  \mathcal{B}_{jk}^{-1} D^2_{kj} \vect{A}_l(q) p_l
    \right) \\
    &\quad - D_i \left( V(q) + \frac{\hbar}{4}(\mathcal{B}_{jk}^{-1} D_{kj}^2V(q)) \right),\\
    \dot{\mathcal{A}}_{ij} &=  - \frac{1}{m}( \mathcal{A}^2 - \mathcal{B}^2)_{ij} + \frac{1}{m} \left( D_i \vect{A}_k(q) \mathcal{A}_{kj}  + \mathcal{A}_{ik} D_j \vect{A}_k(q)  + D_{ij}^2 \vect{A}_k(q) p_k - \frac{1}{2} D_{ij}^2|\vect{A}(q)|^2 \right) - D_{ij}^2V(q),\\
    \dot{\mathcal{B}}_{ij} &=  - \frac{1}{m} (\mathcal{A}\mathcal{B} + \mathcal{B} \mathcal{A})_{ij} + \frac{1}{ m} ( D_{i} \vect{A}_k(q) \mathcal{B}_{kj} + \mathcal{B}_{ik} D_{j} \vect{A}_k(q) ).
  \end{split}
\end{equation}

If we define those terms involving scalar and vector potentials with $O(\hbar)$ corrections as
\begin{gather*}
  V_{\hbar}(q, \mathcal{B}) \defeq V(q) + \frac{\hbar}{4} \Tr(\mathcal{B}^{-1} D^2 V(q) ), \\
  \vect{A}_{\hbar,i}(q, \mathcal{B}) \defeq \vect{A}_i(q) + \frac{\hbar}{4} \Tr(\mathcal{B}^{-1} D^2 \vect{A}_i(q) ),
  \qquad
  |\vect{A}|^{2}_{\hbar}(q, \mathcal{B}) \defeq |\vect{A}(q)|^{2} + \frac{\hbar}{4} \Tr(\mathcal{B}^{-1} D^2 |\vect{A}(q)|^{2} ),
\end{gather*}
we can rewrite the first two of the above set of equations in a slightly more succinct form:
\begin{align*}
  \dot{q} &= \frac{1}{m}\left( p - \vect{A}_{\hbar}(q,\mathcal{B}) \right),\\
  \dot{p} &= -\frac{1}{2m}\grad_{q} \left(
            |\vect{A}|^{2}_{\hbar}(q, \mathcal{B}) - 2 \vect{A}_{\hbar}(q, \mathcal{B}) \cdot p
            \right) - \grad_{q} V_{\hbar}(q, \mathcal{B}).
\end{align*}
Notice also its similarity to the classical equations~\eqref{eq:classical}.

\subsection{Linear Vector Potential with Quadratic Scalar Potential} 
As mentioned in the Introduction, when the vector potential $\vect{A}$ is linear ($\vect{A}(x) = Ax$, where $A$ is a constant $d\times d$ matrix) and the scalar potential $V$ is quadratic, the Gaussian wave packet~\eqref{eq:GWP} gives an exact solution to the Schr\"odinger equation if the parameters satisfy the following set of equations (along with additional equations for $\phi$ and $\delta$):
\begin{equation}
  \label{eq:Zhou}
  \begin{split}
    \dot{q}_i &= \frac{1}{m} (p_{i} - \vect{A}_i(q)),\\
    \dot{p}_i &= -\frac{1}{2m} D_i\left( | \vect{A}(q) |^2 - 2 \vect{A}_j(q) p_j \right) - D_i V(q), \\
    \dot{\mathcal{A}}_{ij} &= \frac{1}{m} ( D_{i} \vect{A}_k(q) \mathcal{A}_{kj}  + \mathcal{A}_{ik} D_j \vect{A}_k(q)) - \frac{1}{2m} D_{ij}^2|\vect{A}(q)|^2 - \frac{1}{m}( \mathcal{A}^2 - \mathcal{B}^2)_{ij} - D_{ij}^2V(q),\\
    \dot{\mathcal{B}}_{ij} &= \frac{1}{ m} ( \mathcal{B}_{ik} D_j \vect{A}_k(q) + D_i \vect{A}_k(q) \mathcal{B}_{kj} ) - \frac{1}{m} (\mathcal{A}\mathcal{B} + \mathcal{B} \mathcal{A})_{ij}.
  \end{split}
\end{equation}
This result is a special case of the more general result of \citet{Ha1998} on quadratic Hamiltonians, and also is equivalent to the set of equations given by \citet{Zh2014}.
We note that both \citeauthor{Ha1998} and \citeauthor{Zh2014} use, instead of $(\mathcal{A},\mathcal{B})$, parameters $(Q,P)$ that are $d\times d$ complex matrices satisfying $Q^{T}P - P^{T}Q = 0$ and $Q^{*}P - P^{*}Q = 2\text{i} I_{d}$; more precisely, \citeauthor{Ha1998} uses parameters $A, B \in \mathbb{C}^{d\times d}$, which are related to $Q$ and $P$ as $A = Q$ and $B = -\text{i} P$.
In fact, these two sets of parameters are related by $\mathcal{A} + \text{i} \mathcal{B} = PQ^{-1}$; see \citet{Oh2015c} for the geometric interpretation of these two different parametrizations.
It is straightforward calculations using this relation to check that \citeauthor{Zh2014}'s equations imply the above set of equations.

Our set of equations, either \eqref{eq:semiclassical-original} or \eqref{eq:semiclassical}, recovers the above set of equations under the above assumptions on the potentials.
In fact, as mentioned above, the asymptotic expansion~\eqref{eq:asymptotic_expansion} is exact if $\mathcal{U}$ is quadratic.
This implies that the Hamiltonian~\eqref{eq:H} and its $\mathcal{O}(\hbar^{2})$ approximation~\eqref{eq:H_hbar} coincide, and thus so do the equations~\eqref{eq:semiclassical-original} and \eqref{eq:semiclassical}.
Now, if the vector potential $\vect{A}$ is linear and the scalar potential $V$ is quadratic, many of the terms in \eqref{eq:semiclassical} involving the Hessians of the potentials vanish, hence recovering \eqref{eq:Zhou}.

\section{Semiclassical Angular Momentum in Electromagnetic Potentials}
\label{sec:angular_momentum}
One advantage of the Hamiltonian formulation using the language of symplectic geometry is that it is amenable to the geometric treatment of symmetry.
Specifically, if the Hamiltonian function of the system is invariant under some Lie group action, it is desirable to investigate any conserved quantities in the system via Noether's Theorem.
Particularly, in this section, we show that the semiclassical angular momentum found in our previous work~\cite{Oh2015b} is conserved if the electromagnetic potentials possess a rotational symmetry.

\subsection{Symmetry in Electromagnetic Potentials}
Suppose that the scalar and vector potentials $V$ and $\vect{A}$ possess the rotational symmetry in the following sense: For any $R \in \mathsf{SO}(d)$ and any $x \in \mathbb{R}^{d}$,
\begin{equation}
  \label{eq:symmetry}
  V(Rx) = V(x)
  \quad\text{and}\quad
  \vect{A}(Rx) = R \vect{A}(x);
\end{equation}
that is, $V$ is $\mathsf{SO}(d)$-invariant whereas $\vect{A}$ is $\mathsf{SO}(d)$-equivariant.
We note that the latter condition implies $D\vect{A}(R x) = R \vect{A}(x) R^{T}$ for any $R \in \mathsf{SO}(d)$ and any $x \in \mathbb{R}^{d}$.

As is done in \cite{Oh2015b}, we define the action of the rotation group $\mathsf{SO}(d)$ on the symplectic manifold $\overline{M} = T^* \mathbb{R}^d \times \Sigma_d$ as follows:
\begin{equation*}
  \Gamma\colon \mathsf{SO}(d)\times \overline{M} \rightarrow \overline{M};
  \qquad
  (q,p,\mathcal{A},\mathcal{B}) \mapsto \Gamma_{R}(q,p,\mathcal{A},\mathcal{B}) \defeq \left(Rq, Rp, R\mathcal{A}R^T, R\mathcal{B}R^T\right).
\end{equation*}
It is easy to check that $\Gamma$ is symplectic, i.e., $\Gamma^*_R \Omega = \Omega$ for any $R \in \mathsf{SO}(d)$.
Then our Hamiltonian, either \eqref{eq:H} or \eqref{eq:H_hbar}, is invariant under this action, i.e., for any $R \in \mathsf{SO}(d)$, $H \circ \Gamma_{R} = H$ and $H_{\hbar} \circ \Gamma_{R} = H_{\hbar}$.
In fact, for the Hamiltonian~\eqref{eq:H_hbar}, it follows from a straightforward calculation using the above symmetry assumptions on $V$ and $\vect{A}$.
For the Hamiltonian~\eqref{eq:H}, note that the expectation values of the potentials maintain the same symmetry, i.e.,
\begin{align*}
  \left\langle V\right\rangle(Rq, R\mathcal{B}R^T) = \left\langle V\right\rangle(q,\mathcal{B}),
  \quad\text{and}\quad
  \left\langle \vect{A}\right\rangle(Rq, R\mathcal{B}R^T) = R \left\langle \vect{A}\right\rangle(q,\mathcal{B}).
\end{align*}

\subsection{Semiclassical Angular Momentum}
The momentum map $\mathbf{J}_{\hbar}\colon \overline{M} \to \mathfrak{so}(d)^{*}$ corresponding to the action $\Gamma$ defined above is given by~(see \citet[Section 3]{Oh2015b} for the derivation)
\begin{equation}
  \label{eq:J}
  \mathbf{J}_{\hbar}(q,p,\mathcal{A},\mathcal{B}) = q \diamond p - \frac{\hbar}{2} [\mathcal{B}^{-1}, \mathcal{A}],
\end{equation}
where $(q \diamond p)_{ij} = q_j p_i - q_i p_j$ (see \citet[Remark 6.3.3]{Ho2011b}), and we identified $\mathfrak{so}(d)^{*}$ with $\mathfrak{so}(d)$ via an inner product.
Setting $\hbar = 0$ reduces the above to the classical angular momentum, hence we call the above the \textit{semiclassical angular momentum}.
Interestingly, this semiclassical angular momentum coincides with the expectation value of the angular momentum with respect to the normalized Gaussian, i.e., for $d = 3$,
\begin{equation*}
  \left\langle \hat{x} \times \hat{p} \right\rangle
  = \mathbf{J}_{\hbar}(q,p,\mathcal{A},\mathcal{B}).
\end{equation*}

Now, assuming the symmetry~\eqref{eq:symmetry} in the potentials, by Noether's Theorem (see, e.g., \citet[Theorem 11.4.1]{MaRa1999}), we conclude that the semiclassical angular momentum~\eqref{eq:J} is a conserved quantity of our semiclassical equation~\eqref{eq:semiclassical-original} or \eqref{eq:semiclassical}.

\section{Numerical Examples}
Given that our set of equations~\eqref{eq:semiclassical} differs from that of \citeauthor{Zh2014} by $\mathcal{O}(\hbar)$ correction terms, a natural question is whether these correction terms improve the accuracy of approximation.
Specifically, we are interested in comparing the time evolution $t \mapsto z(t) = (q(t), p(t))$ of the phase space variables of our semiclassical equations with that of the expectation values $\langle \hat{z}\rangle$ of the position and momentum operators $\hat{z} = (\hat{x}, \hat{p})$, i.e., $t \mapsto \langle \hat{z}\rangle(t) = \langle \psi(t,\,\cdot\,), \hat{z} \psi(t,\,\cdot\,)\rangle$, where $t \mapsto \psi(t,\,\cdot\,)$ is a solution of the Schr\"odinger equation~\eqref{eq:Schroedinger}.

In the following, we compare numerical solutions of the classical equations~\eqref{eq:classical}, the semiclassical equations~\eqref{eq:semiclassical}, as well as the time-dependent expectation values of observables as calculated by the Egorov \cite{Eg1969,CoRo2012,LaRo2010} or Initial Value Representation~\cite{Mi1970,Mi1974b,WaSuMi1998,Mi2001} (Egorov/IVR) method.
Note that the time evolution of $(q,p)$ of \citeauthor{Zh2014}'s equations~\eqref{eq:Zhou} is identical to that of the classical equations~\eqref{eq:classical}.

We employ the Egorov/IVR method because it is computationally prohibitive to solve for the highly oscillatory wave functions numerically in the semiclassical regime.
It is also suited for our purposes because we are interested in the time evolution of expectation values.
In the Egorov/IVR method, the Wigner function of the initial wave function is calculated. An observable is evaluated along the solutions of the classical system for each sampled point in phase space where the phase space is sampled according to the Wigner function. This gives an $\mathcal{O}(\hbar^2)$ approximation of the expectation value of that observable, with an error proportional to $1/\sqrt{N}$, where $N$ is the number of samples.

In all of the following, we solved our equations and the classical equations by the explicit Runge--Kutta method with a time step of $0.01$. For the Egorov/IVR computations, we used $10^6$ samples for each value of $\hbar$, with the exception of $\hbar = 0.01$ for which we used $10^7$ samples.

\subsection{1D Example}
Here we let $d=1$, $m = 1$, $V(x) = 1 - \frac{1}{2} \cos^2(x)$, $A(x) = \cos(x)$, subject to the initial conditions $q(0) = 0.5$, $p(0) = -1$, $\mathcal{A}(0) = 0$, $\mathcal{B}(0) = 1$; the scalar and vector potentials are taken from \citet[Example~1]{Zh2014}.
In order to see how the error converges as $\hbar \to 0$, we ran the computations for $\hbar = 0.5, 0.3, 0.1, 0.05, 0.03, 0.01$.

\begin{figure}[htbp]
\centering
\subfloat[$\hbar = 0.5$]{
  \includegraphics[width=78mm]{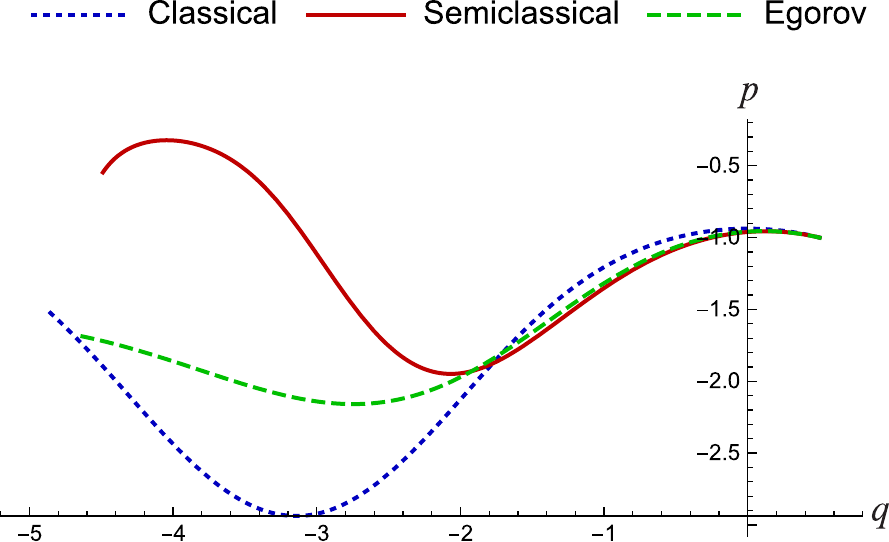}
}
\subfloat[$\hbar = 0.1$]{
  \includegraphics[width=78mm]{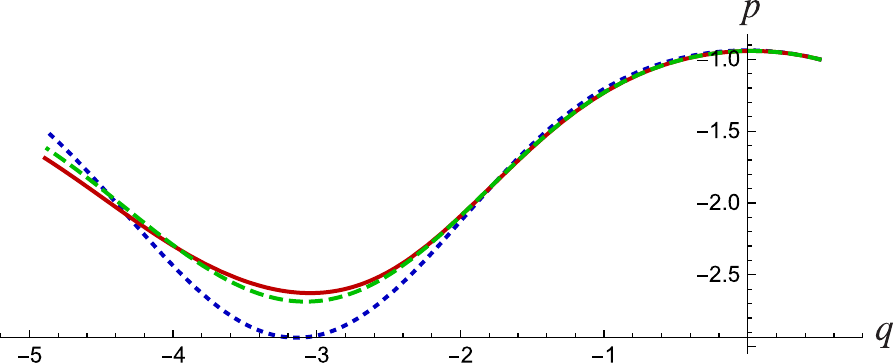}
}
\hspace{0mm}
\subfloat[$\hbar = 0.05$]{
  \includegraphics[width=78mm]{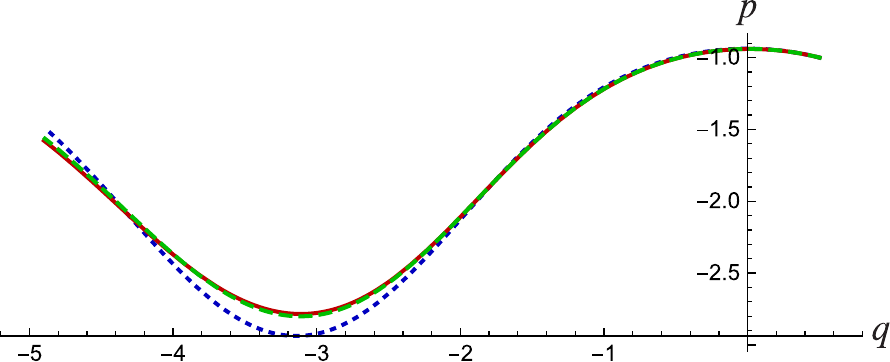}
}
\subfloat[$\hbar = 0.01$]{
  \includegraphics[width=78mm]{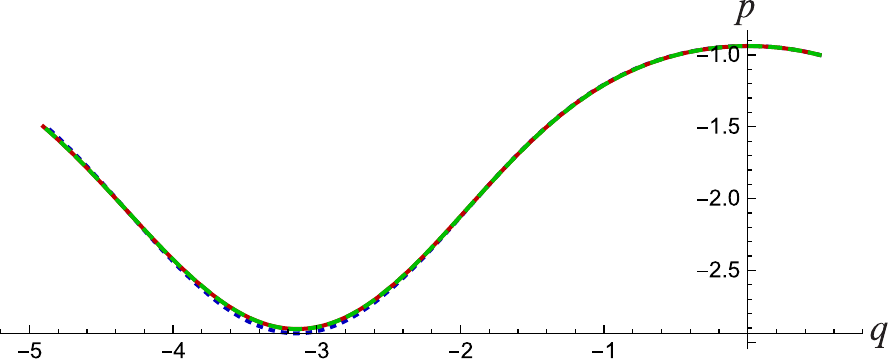}
}\\
\subfloat[Convergence of errors as $\hbar \to 0$]{
  \includegraphics[width=85mm]{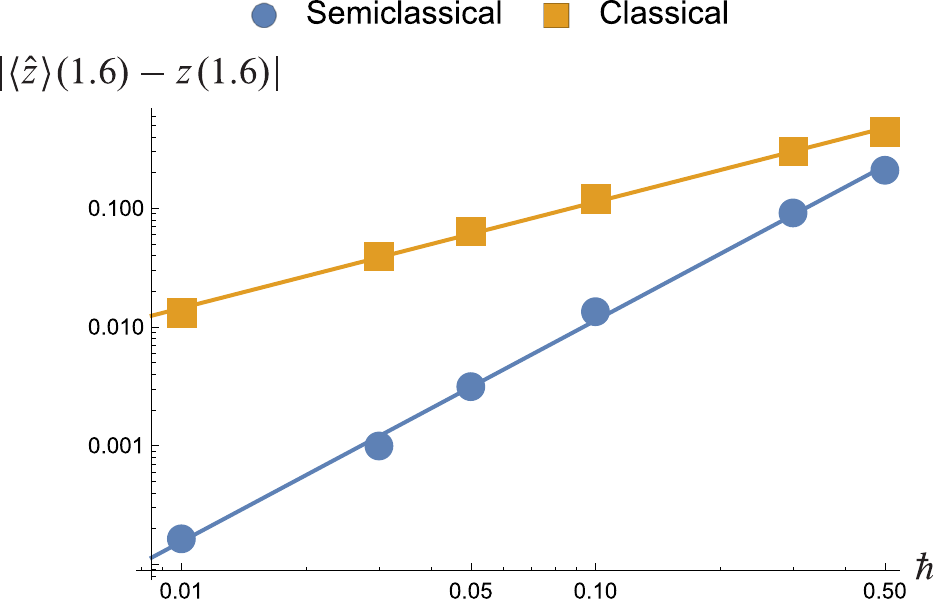}
}
\caption{
  Results of 1D computations with $m = 1$, $V(x) = 1 - \frac{1}{2} \cos^2(x)$, $A(x) = \cos(x)$.
  (a)--(d): Parametric plots of $t \mapsto q(t) = (q_{1}(t), q_{2}(t))$ in the classical phase space $T^{*}\mathbb{R} \cong \mathbb{R}^{2}$ for $\hbar = 0.5, 0.1, 0.05, 0.01$ from $t=0$ to $t=3$.
  Our solutions are closer to the Egorov/IVR than the classical solutions.
  (e): The error $|\langle\hat{z}\rangle(t) - z(t)|$ for several values of $\hbar$ at $t = 1.6$.
  As $\hbar \rightarrow 0$, our solutions converge to the Egorov/IVR solutions faster than the classical equations.
  The equation of the best fit line for the semiclassical error is $\exp(-0.190) * \hbar^{1.864}$, and $\exp(-0.125)*\hbar^{0.894}$ for the classical.
}
\label{fig:1d}
\end{figure}

Figure~\ref{fig:1d} shows the solutions on the classical phase space $T^{*}\mathbb{R} = \mathbb{R}^{2}$ from $t=0$ to $t=3$ as well as the error $|\langle\hat{z}\rangle(t) - z(t)|$ at $t = 1.6$ in terms of the Euclidean norm on the classical phase space.
As can be seen, our solutions are closer to the Egorov/IVR than the classical solutions. Furthermore, as $\hbar \rightarrow 0$, our solutions converge to the Egorov/IVR solutions faster than the classical equations.

\begin{figure}[htbp]
\centering
\subfloat[$\hbar = 0.5$]{
  \includegraphics[width=80mm]{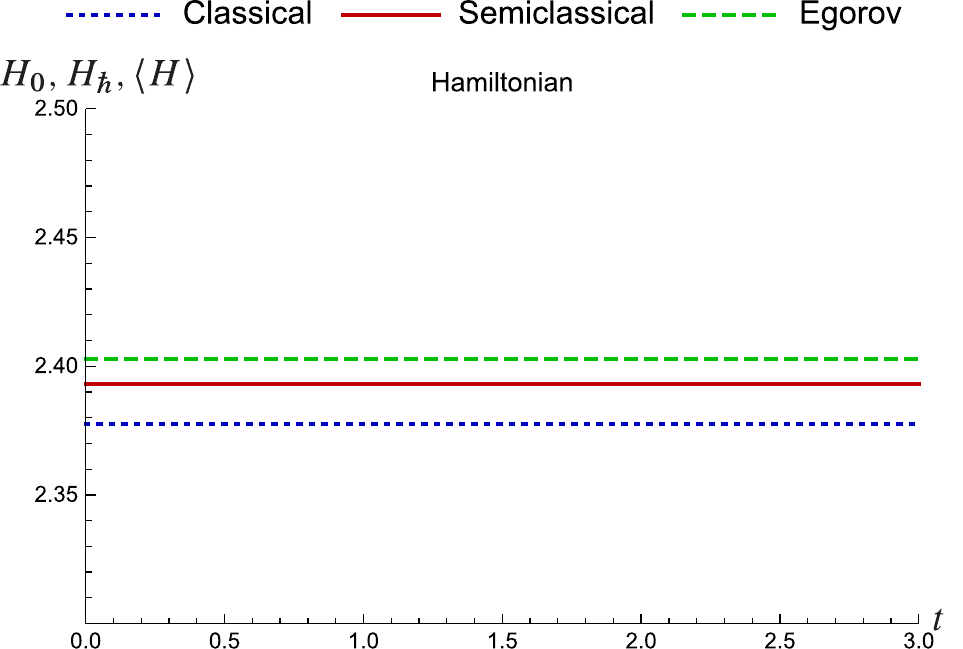}
}
\subfloat[$\hbar = 0.1$]{
  \includegraphics[width=80mm]{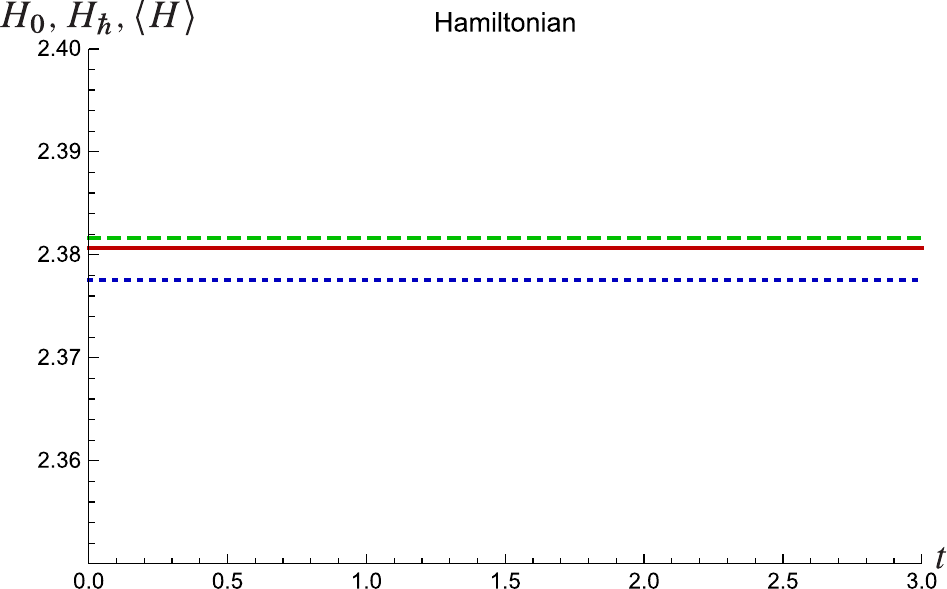}
}
\hspace{0mm}
\subfloat[$\hbar = 0.05$]{
  \includegraphics[width=80mm]{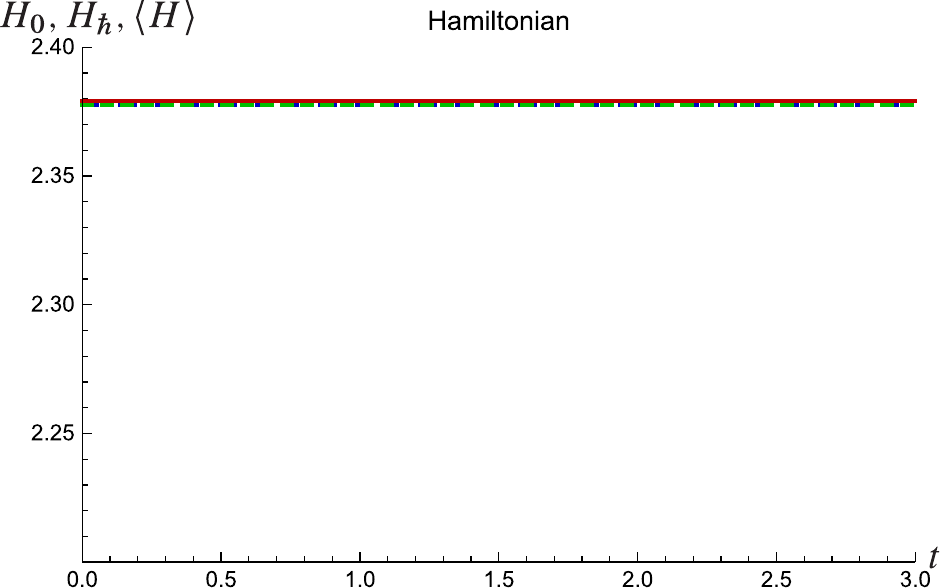}
}
\subfloat[$\hbar = 0.01$]{
  \includegraphics[width=80mm]{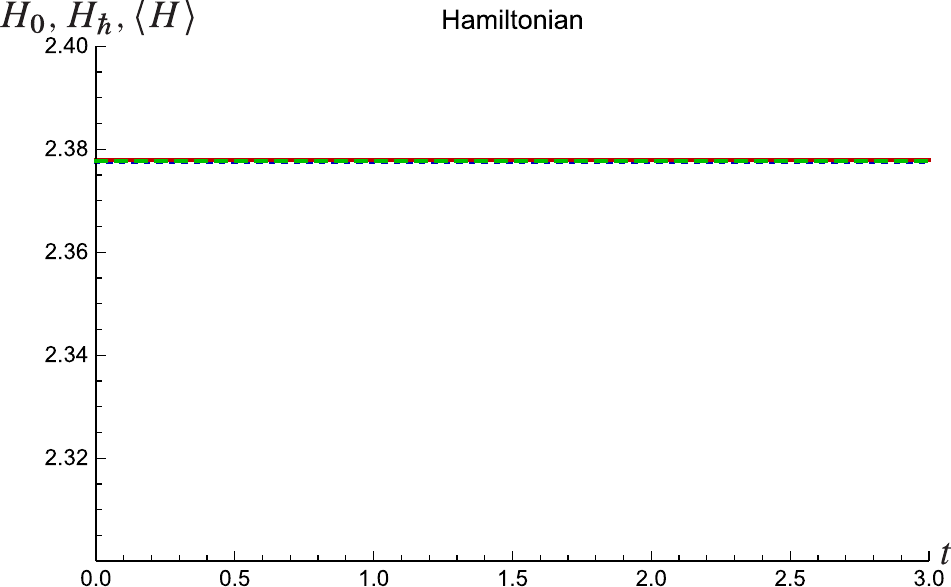}
}
\caption{Time evolution of the Hamiltonian for the above 1D system solutions for $\hbar = 0.5, 0.1, 0.05, 0.01$. The semiclassical Hamiltonian~\eqref{eq:H_hbar} more closely approximates the Egorov/IVR expectation value $\langle\hat{H}\rangle$ of the Hamiltonian operator than the classical Hamiltonian~\eqref{eq:H_0}.}
\label{fig:H-1d}
\end{figure}

Figure~\ref{fig:H-1d} shows the time evolutions of the Hamiltonians for the classical, semiclassical, and Egorov/IVR solutions.
Note that the Hamiltonians for all these three cases are different:
It is $H_{0}$ in \eqref{eq:H_0} for the classical case and $H_{\hbar}$ in \eqref{eq:H_hbar} for the semiclassical case, whereas for the Egorov/IVR case, it is the expectation value $\langle \hat{H}\rangle$ of the Hamiltonian operator $\hat{H}$.
In each of these cases, the corresponding Hamiltonian is a conserved quantity.
Notice that the semiclassical Hamiltonian gives a better approximation to the expectation value of the Hamiltonian.

\subsection{2D Example}
Here we let $d=2$, $V(x) = \frac{1}{2} |x|^2 + \frac{1}{4} |x|^4 $, $A(x) = (-x_2, x_1)$, subject to the initial conditions $q(0) = (1,0)$, $p(0) = (0,1)$, $\mathcal{A}(0) = \begin{pmatrix} -3 & -6 \\ -6 & -6 \end{pmatrix}$, $\mathcal{B}(0) = \begin{pmatrix} 1 & 1/2 \\ 1/2 & 1 \end{pmatrix}$.

\begin{figure}[ht]
\centering
\subfloat[$\hbar = 0.5$]{
  \includegraphics[width=78mm]{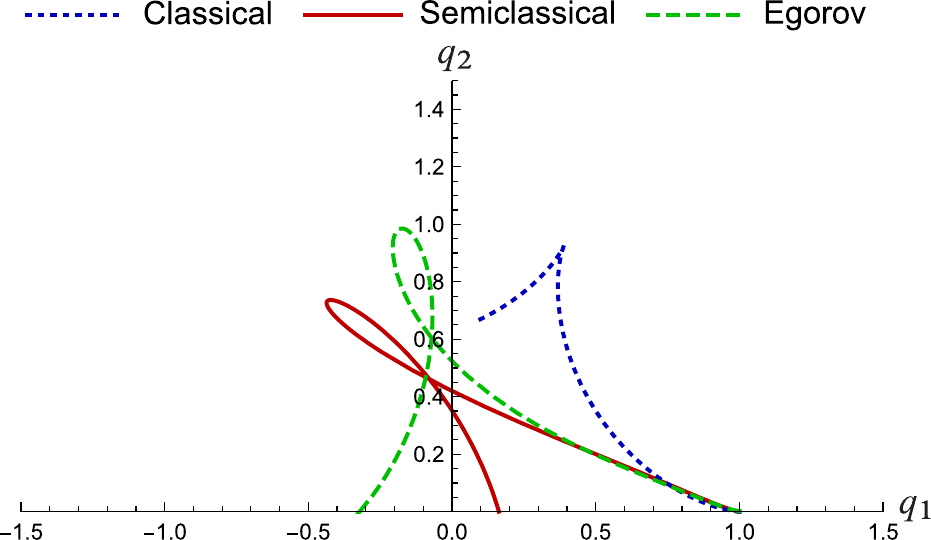}
}
\subfloat[$\hbar = 0.1$]{
  \includegraphics[width=78mm]{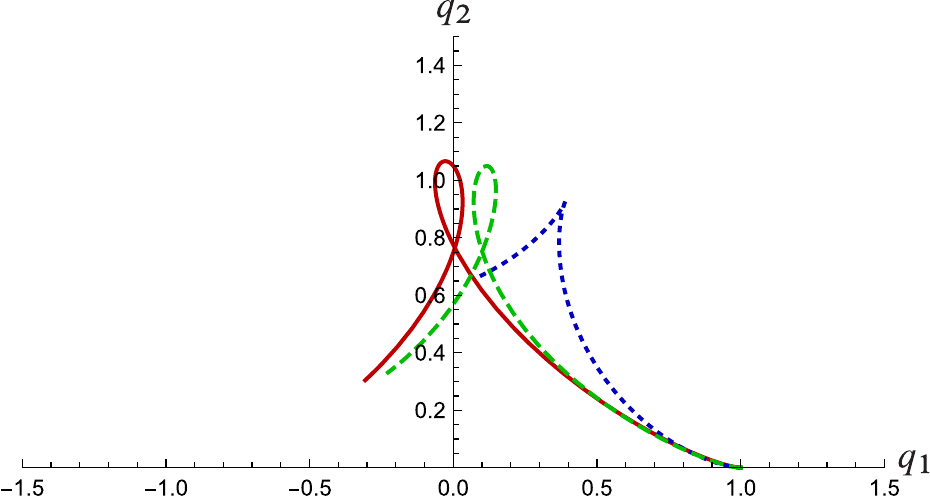}
}
\hspace{0mm}
\subfloat[$\hbar = 0.05$]{
  \includegraphics[width=78mm]{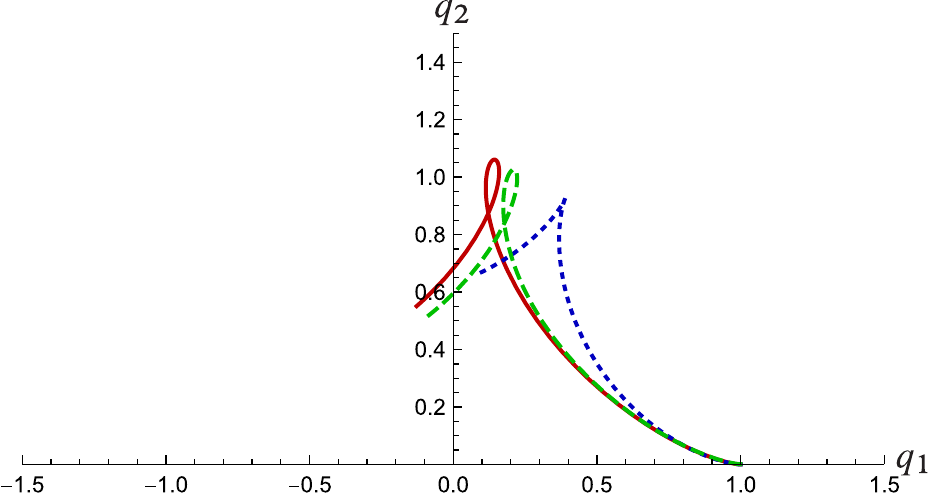}
}
\subfloat[$\hbar = 0.01$]{
  \includegraphics[width=78mm]{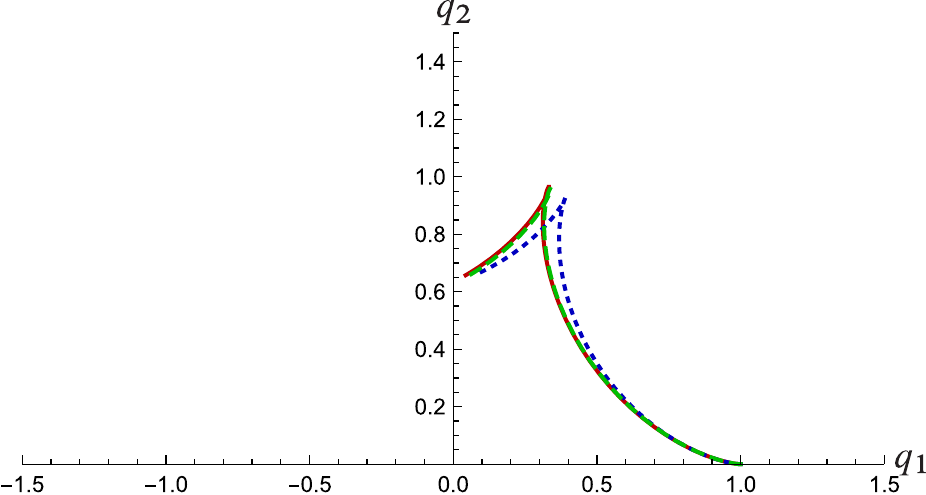}
}\\
\subfloat[Convergence of errors as $\hbar \to 0$]{
  \includegraphics[width=85mm]{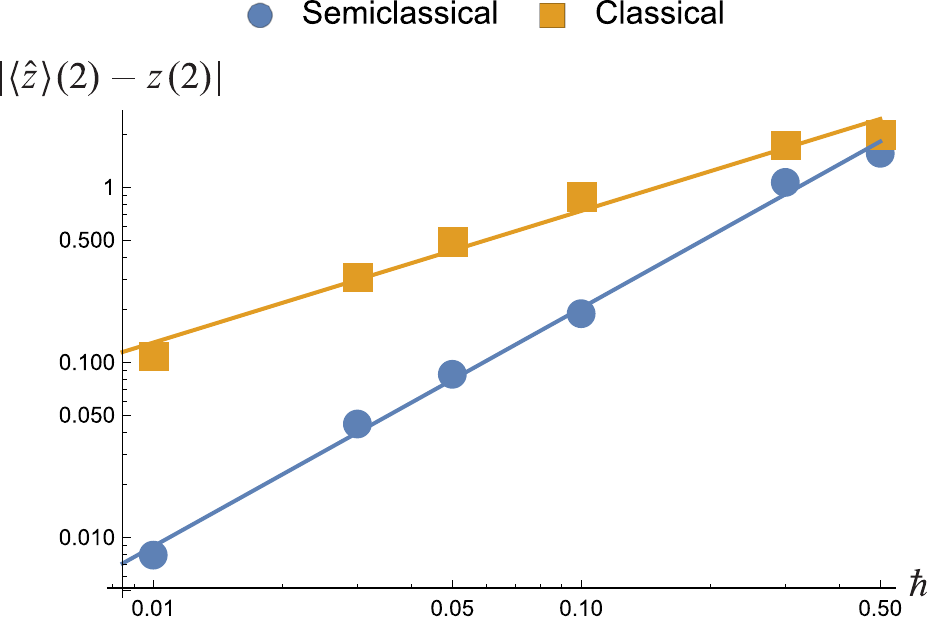}
}
\caption{
  Results of 2D computations with $d=2$, $V(x) = \frac{1}{2} |x|^2 + \frac{1}{4} |x|^4 $, $A(x) = (-x_2, x_1)$.
  (a)--(d): Parametric plots of $t \mapsto q(t) = (q_{1}(t), q_{2}(t))$ in the classical configuration space $\mathbb{R}^{2}$ for $\hbar = 0.5, 0.1, 0.05, 0.01$ from $t=0$ to $t=3$.
  (e): The error $|\langle\hat{z}\rangle(t) - z(t)|$ for several values of $\hbar$ at $t = 2$.
  Again, as $\hbar \rightarrow 0$, our solutions converge to the Egorov/IVR solutions faster than the classical equations.
  The equation of the best fit line for the semiclassical error is $\exp (1.544) \hbar^{1.3612}$, and $\exp (1.41845) \hbar^{0.752}$ for the classical.
}
\label{fig:2d}
\end{figure}

\begin{figure}[ht]
\centering
\subfloat[$\hbar = 0.5$]{
  \includegraphics[width=80mm]{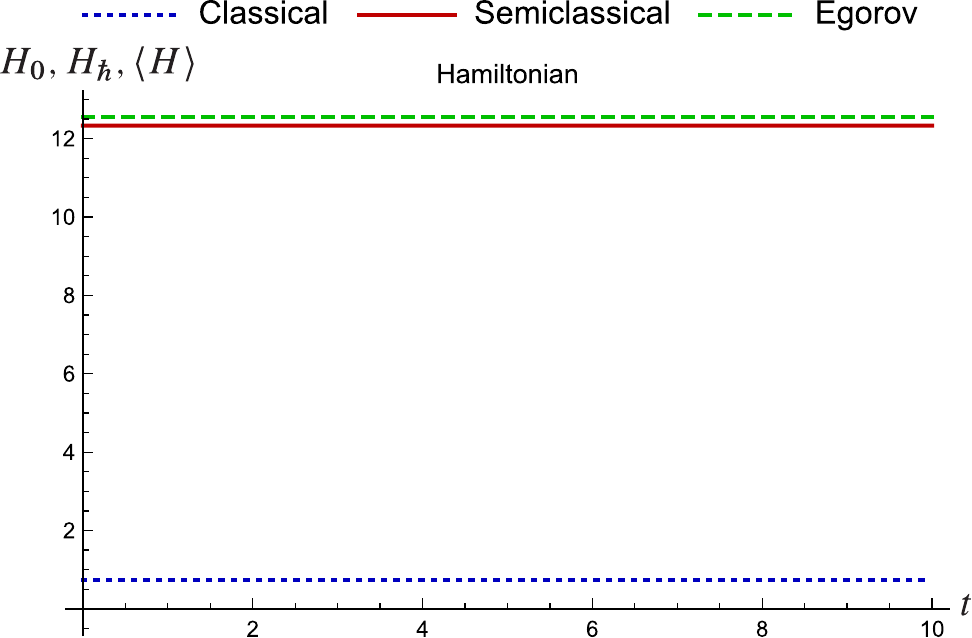}
}
\subfloat[$\hbar = 0.1$]{
  \includegraphics[width=80mm]{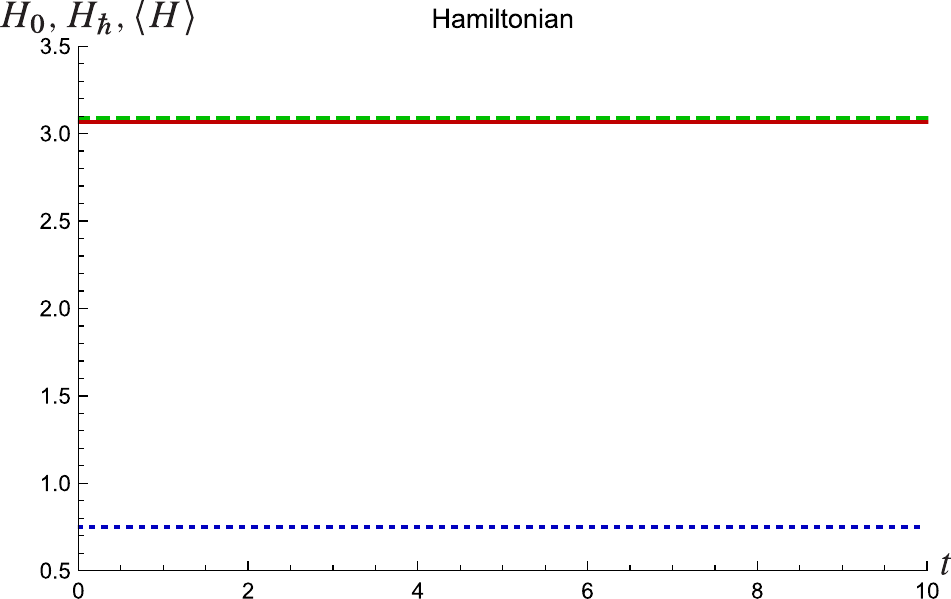}
}
\hspace{0mm}
\subfloat[$\hbar = 0.05$]{
  \includegraphics[width=80mm]{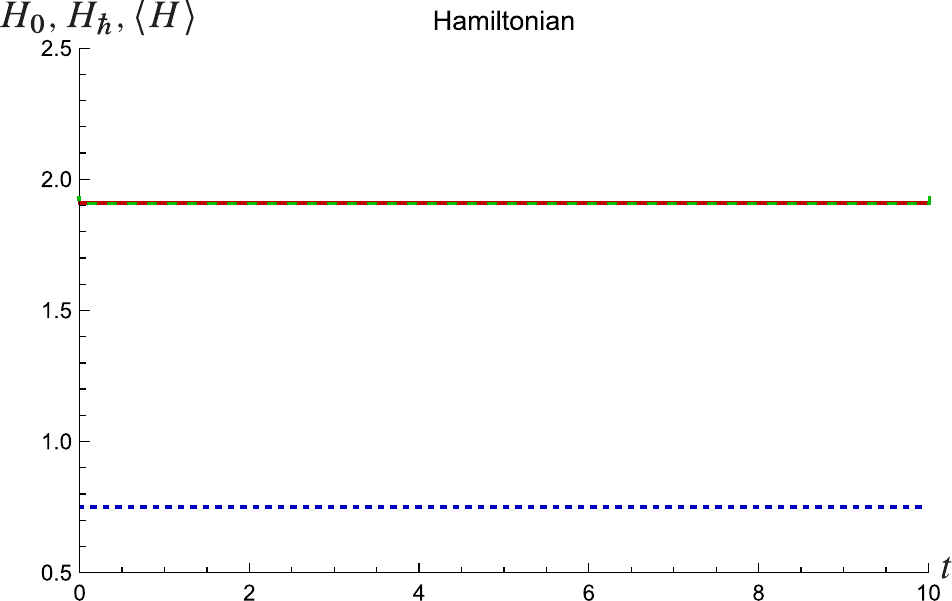}
}
\subfloat[$\hbar = 0.01$]{
  \includegraphics[width=80mm]{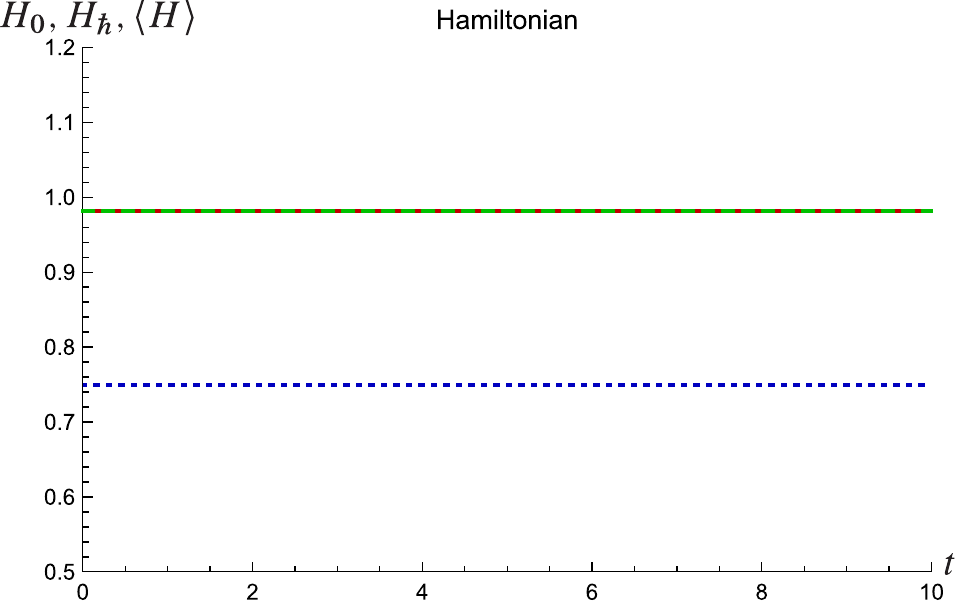}
}
\caption{Time evolution of the Hamiltonian for the above 2D system solutions for $\hbar = 0.5, 0.1, 0.05, 0.01$. The semiclassical Hamiltonian~\eqref{eq:H_hbar} more closely approximates the Egorov/IVR expectation value $\langle\hat{H}\rangle$ of the Hamiltonian operator than the classical Hamiltonian~\eqref{eq:H_0}.}
\label{fig:H-2d}
\end{figure}

Figure~\ref{fig:2d} shows the solutions on the classical configuration space $\mathbb{R}^{2} = \{(q_{1}, q_{2})\}$ from $t=0$ to $t=10$ as well as the error $|\langle\hat{z}\rangle(t) - z(t)|$ at $t = 2$ in terms of the Euclidean norm on the classical phase space $T^{*}\mathbb{R}^{2} \cong \mathbb{R}^{4}$.
Figure~\ref{fig:H-2d} shows the time evolutions of the Hamiltonians for the classical, semiclassical, and Egorov/IVR solutions just as in the 1D case.
The same observations as above apply to these 2D results as well.

\begin{figure}[ht]
\centering
\subfloat[$\hbar = 0.5$]{
  \includegraphics[width=80mm]{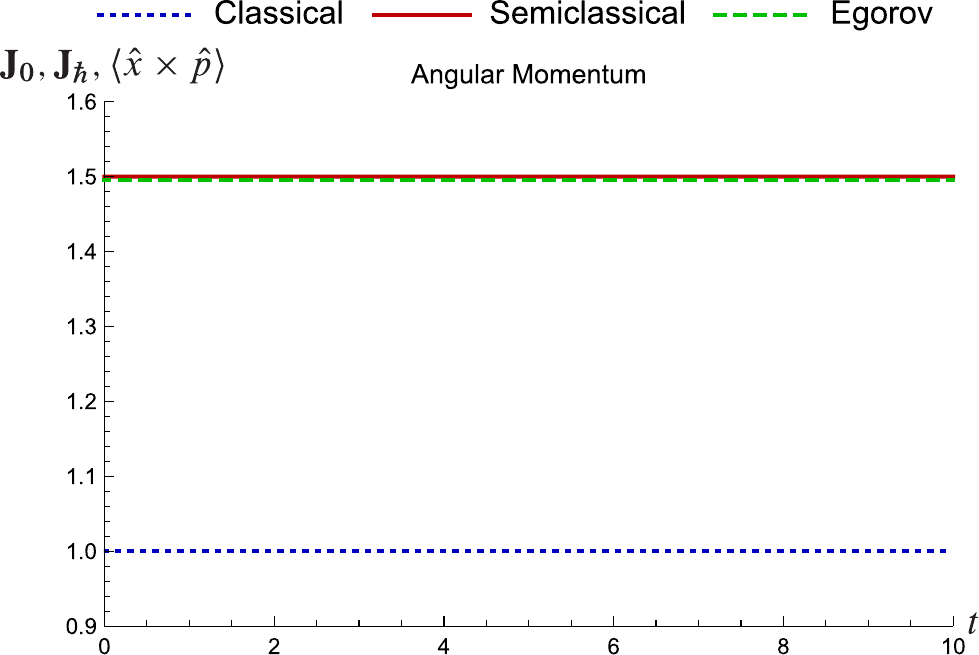}
}
\subfloat[$\hbar = 0.1$]{
  \includegraphics[width=80mm]{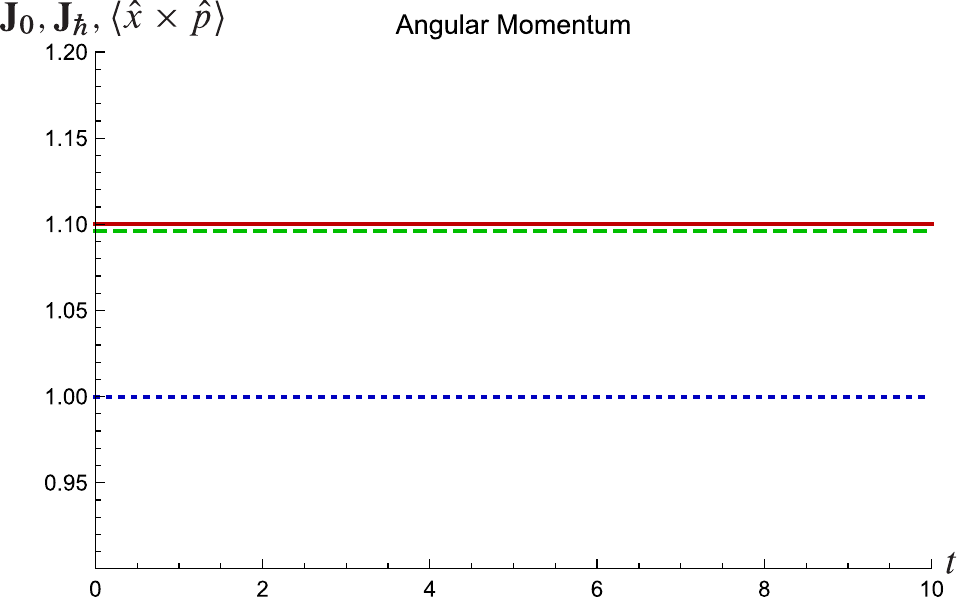}
}
\hspace{0mm}
\subfloat[$\hbar = 0.05$]{
  \includegraphics[width=80mm]{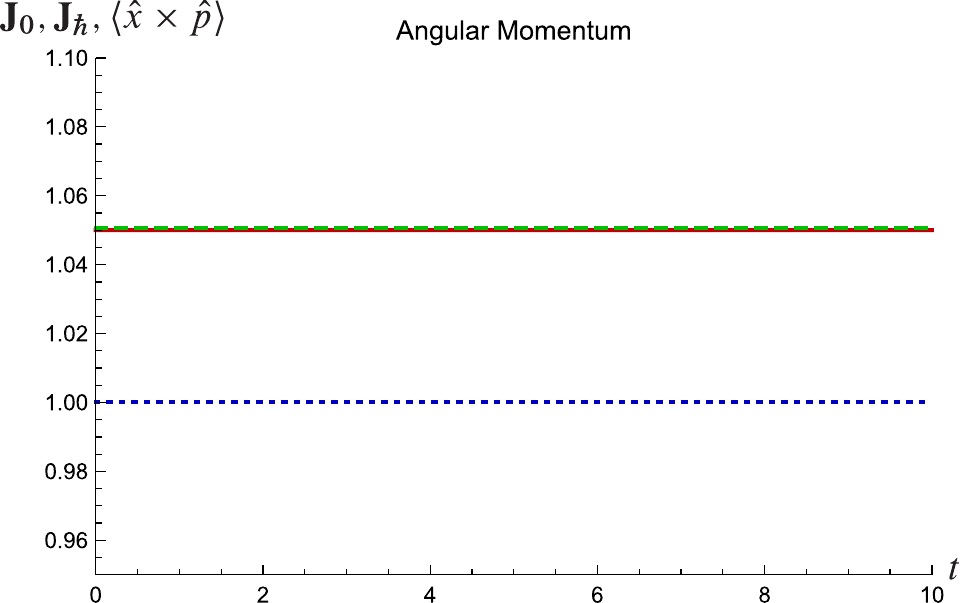}
}
\subfloat[$\hbar = 0.01$]{
  \includegraphics[width=80mm]{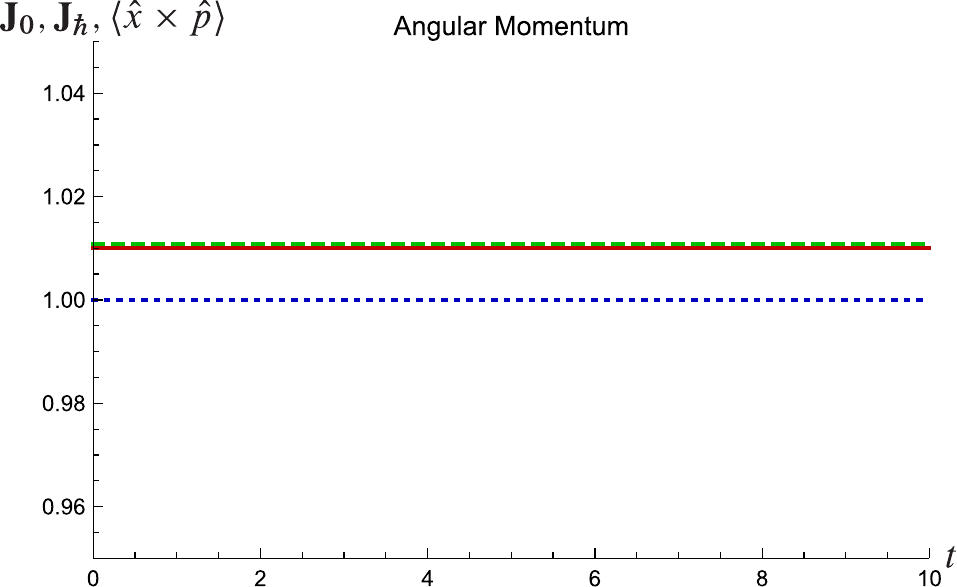}
}
\caption{Time evolution of the classical angular momentum along the classical solution, the semiclassical angular momentum along the semiclassical solution, and the expectation value of the angular momentum operator for the 2D system. The semiclassical angular momentum is in far closer agreement than the angular momentum along the classical solutions.}
\label{fig:angular_momentum}
\end{figure}

For this 2D example, the scalar and vector potentials chosen above satisfy the symmetry condition~\eqref{eq:symmetry}.
Therefore, based on the result of Section~\ref{sec:angular_momentum}, the semiclassical angular momentum~\eqref{eq:J} is also a conserved quantity of the semiclassical system~\eqref{eq:semiclassical} as well.
Figure~\ref{fig:angular_momentum} shows the time evolutions of the classical angular momentum along the classical solutions, the semiclassical angular momentum~\eqref{eq:J} along the semiclassical solutions, and the expectation value of the angular momentum operator along the Egorov/IVR solutions.
We see that the semiclassical angular momentum gives a better approximation to the expectation value of the angular momentum than the classical one does.

\section{Conclusion and Future Work}
We extended our earlier work on the Hamiltonian formulation of Gaussian wave packets to incorporate electromagnetic fields.
Many of the results are extensions of our previous works to incorporate the electromagnetic effects.
These results greatly expand the range of applications of semiclassical dynamics because of its importance in quantum control and solid state physics.

As seen in the above numerical results, our solutions converge to the the expectation value of the operator $z = (q,p)$ along the Egorov/IVR solution faster than the classical solution. Since the equations for $q$ and $p$ given by \citet{Zh2014} are identical to the classical equations, our solutions also converge faster than those of \citeauthor{Zh2014}. These results demonstrate that the $\mathcal{O}(\hbar)$ correction terms in our semiclassical equations~\eqref{eq:semiclassical} indeed improve the accuracy of the approximations of expectation values.

Our preliminary studies (under certain technical assumptions and without electromagnetic fields) indicate that the errors in the observables of the classical solution is $\mathcal{O}(\hbar)$ whereas $\mathcal{O}(\hbar^{3/2})$ for the semiclassical solution, despite the well-known fact that the Gaussian wave packet dynamics gives $\mathcal{O}(\hbar^{1/2})$ approximation \textit{in terms of the wave functions} in $L^{2}$-norm established by \citet{Ha1980, Ha1981, Hagedorn1985, Ha1998}.
Our numerical results seem to support these claims.
A proof of this error estimate remains for a future work.

\bibliography{main}

\begin{thebibliography}{26}
\providecommand{\natexlab}[1]{#1}
\providecommand{\url}[1]{\texttt{#1}}
\expandafter\ifx\csname urlstyle\endcsname\relax
  \providecommand{\doi}[1]{doi: #1}\else
  \providecommand{\doi}{doi: \begingroup \urlstyle{rm}\Url}\fi

\bibitem[Combescure and Robert(2012)]{CoRo2012}
M.~Combescure and D.~Robert.
\newblock \emph{Coherent States and Applications in Mathematical Physics}.
\newblock Springer, 2012.

\bibitem[Egorov(1969)]{Eg1969}
Y.~V. Egorov.
\newblock The canonical transformations of pseudodifferential operators.
\newblock \emph{Uspekhi Mat. Nauk}, 24\penalty0 (5(149)):\penalty0 235--236,
  1969.

\bibitem[Hagedorn(1980)]{Ha1980}
G.~A. Hagedorn.
\newblock Semiclassical quantum mechanics. {I}. {T}he $\hbar\to0$ limit for
  coherent states.
\newblock \emph{Communications in Mathematical Physics}, 71\penalty0
  (1):\penalty0 77--93, 1980.

\bibitem[Hagedorn(1981)]{Ha1981}
G.~A. Hagedorn.
\newblock Semiclassical quantum mechanics. {III}. the large order asymptotics
  and more general states.
\newblock \emph{Annals of Physics}, 135\penalty0 (1):\penalty0 58--70, 1981.

\bibitem[Hagedorn(1985)]{Hagedorn1985}
G.~A. Hagedorn.
\newblock Semiclassical quantum mechanics, {IV}: large order asymptotics and
  more general states in more than one dimension.
\newblock \emph{Annales de l'institut Henri Poincar{\'e} (A) Physique
  th{\'e}orique}, 42\penalty0 (4):\penalty0 363--374, 1985.

\bibitem[Hagedorn(1998)]{Ha1998}
G.~A. Hagedorn.
\newblock Raising and lowering operators for semiclassical wave packets.
\newblock \emph{Annals of Physics}, 269\penalty0 (1):\penalty0 77--104, 1998.

\bibitem[Heller(1975)]{He1975a}
E.~J. Heller.
\newblock Time-dependent approach to semiclassical dynamics.
\newblock \emph{Journal of Chemical Physics}, 62\penalty0 (4):\penalty0
  1544--1555, 1975.

\bibitem[Heller(1976)]{He1976b}
E.~J. Heller.
\newblock Classical {$S$}-matrix limit of wave packet dynamics.
\newblock \emph{Journal of Chemical Physics}, 65\penalty0 (11):\penalty0
  4979--4989, 1976.

\bibitem[Heller(1981)]{He1981}
E.~J. Heller.
\newblock Frozen {G}aussians: A very simple semiclassical approximation.
\newblock \emph{Journal of Chemical Physics}, 75\penalty0 (6):\penalty0
  2923--2931, 1981.

\bibitem[Holm(2011)]{Ho2011b}
D.~D. Holm.
\newblock \emph{Geometric Mechanics, Part II: Rotating, Translating and
  Rolling}.
\newblock Imperial College Press, 2nd edition, 2011.

\bibitem[Kramer and Saraceno(1981)]{KrSa1981}
P.~Kramer and M.~Saraceno.
\newblock \emph{Geometry of the time-dependent variational principle in quantum
  mechanics}.
\newblock Lecture notes in physics. Springer-Verlag, 1981.

\bibitem[Lasser and R{\"o}blitz(2010)]{LaRo2010}
C.~Lasser and S.~R{\"o}blitz.
\newblock Computing expectation values for molecular quantum dynamics.
\newblock \emph{SIAM Journal on Scientific Computing}, 32\penalty0
  (3):\penalty0 1465--1483, 2010.

\bibitem[Littlejohn(1986)]{Li1986}
R.~G. Littlejohn.
\newblock The semiclassical evolution of wave packets.
\newblock \emph{Physics Reports}, 138\penalty0 (4-5):\penalty0 193--291, 1986.

\bibitem[Lubich(2008)]{Lu2008}
C.~Lubich.
\newblock \emph{From quantum to classical molecular dynamics: reduced models
  and numerical analysis}.
\newblock European Mathematical Society, Z{\"u}rich, Switzerland, 2008.

\bibitem[Marsden and Ratiu(1999)]{MaRa1999}
J.~E. Marsden and T.~S. Ratiu.
\newblock \emph{Introduction to Mechanics and Symmetry}.
\newblock Springer, 1999.

\bibitem[Marsden and Weinstein(1974)]{MaWe1974}
J.~E. Marsden and A.~Weinstein.
\newblock Reduction of symplectic manifolds with symmetry.
\newblock \emph{Reports on Mathematical Physics}, 5\penalty0 (1):\penalty0
  121--130, 1974.

\bibitem[Marsden et~al.(2007)Marsden, Misiolek, Ortega, Perlmutter, and
  Ratiu]{MaMiOrPeRa2007}
J.~E. Marsden, G.~Misiolek, J.~P. Ortega, M.~Perlmutter, and T.~S. Ratiu.
\newblock \emph{Hamiltonian Reduction by Stages}.
\newblock Springer, 2007.

\bibitem[Miller(2006)]{Mi2006}
P.~D. Miller.
\newblock \emph{Applied Asymptotic Analysis}.
\newblock American Mathematical Society, Providence, R.I., 2006.

\bibitem[Miller(1970)]{Mi1970}
W.~H. Miller.
\newblock Classical {S} matrix: Numerical application to inelastic collisions.
\newblock \emph{The Journal of Chemical Physics}, 53\penalty0 (9):\penalty0
  3578--3587, 1970.

\bibitem[Miller(1974)]{Mi1974b}
W.~H. Miller.
\newblock Quantum mechanical transition state theory and a new semiclassical
  model for reaction rate constants.
\newblock \emph{The Journal of Chemical Physics}, 61\penalty0 (5):\penalty0
  1823--1834, 1974.

\bibitem[Miller(2001)]{Mi2001}
W.~H. Miller.
\newblock The semiclassical initial value representation: A potentially
  practical way for adding quantum effects to classical molecular dynamics
  simulations.
\newblock \emph{The Journal of Physical Chemistry A}, 105\penalty0
  (13):\penalty0 2942--2955, 2001.

\bibitem[Ohsawa(2015{\natexlab{a}})]{Oh2015b}
T.~Ohsawa.
\newblock Symmetry and conservation laws in semiclassical wave packet dynamics.
\newblock \emph{Journal of Mathematical Physics}, 56\penalty0 (3):\penalty0
  032103, 2015{\natexlab{a}}.

\bibitem[Ohsawa(2015{\natexlab{b}})]{Oh2015c}
T.~Ohsawa.
\newblock The {S}iegel upper half space is a {M}arsden--{W}einstein quotient:
  Symplectic reduction and {G}aussian wave packets.
\newblock \emph{Letters in Mathematical Physics}, 105\penalty0 (9):\penalty0
  1301--1320, 2015{\natexlab{b}}.

\bibitem[Ohsawa and Leok(2013)]{OhLe2013}
T.~Ohsawa and M.~Leok.
\newblock Symplectic semiclassical wave packet dynamics.
\newblock \emph{Journal of Physics A: Mathematical and Theoretical},
  46\penalty0 (40):\penalty0 405201, 2013.

\bibitem[Wang et~al.(1998)Wang, Sun, and Miller]{WaSuMi1998}
H.~Wang, X.~Sun, and W.~H. Miller.
\newblock Semiclassical approximations for the calculation of thermal rate
  constants for chemical reactions in complex molecular systems.
\newblock \emph{The Journal of Chemical Physics}, 108\penalty0 (23):\penalty0
  9726--9736, 1998.

\bibitem[Zhou(2014)]{Zh2014}
Z.~Zhou.
\newblock Numerical approximation of the {S}chr{\"o}dinger equation with the
  electromagnetic field by the {H}agedorn wave packets.
\newblock \emph{Journal of Computational Physics}, 272:\penalty0 386--407,
  2014.

\end{thebibliography}
\bibliographystyle{plainnat}
 
\end{document}